\documentclass{IEEEtran}
\usepackage{amsmath}
\usepackage{nccmath}
\usepackage[utf8]{inputenc} 
\usepackage{epsfig,scalefnt,multirow}
\usepackage{url}
\usepackage{ifthen}
\usepackage{mathtools}
\usepackage{cite}
\usepackage{graphicx}
\usepackage{amssymb}
\usepackage{tabularx}
\usepackage{amsmath}
\usepackage{epstopdf}
\usepackage{epsf}
\usepackage{algorithm}
\usepackage{algpseudocode}
\usepackage{algpascal}
\usepackage{cases}
\usepackage{caption}
\usepackage[labelformat=simple]{subcaption}

\usepackage{stfloats}
\usepackage{float}
\usepackage{xcolor}
\usepackage{tabularx}

\usepackage{epsfig,amsmath,amssymb,epsf,amsthm,scalefnt,multirow}
\usepackage{xcolor}
\usepackage{float}
\usepackage{cite}
\usepackage{psfrag}
\usepackage{gensymb}
\usepackage{cleveref}
\usepackage{mathrsfs}
\usepackage{mathtools}
\usepackage{algpseudocode}
\usepackage{algorithm}
\usepackage{enumerate}

\def\b0{{\pmb{0}}} 

  \def\bg{{\mathbf{g}}} \def\bh{{\mathbf{h}}}

\def\by{{\mathbf{y}}} \def\bz{{\mathbf{z}}}

\def\bI{{\mathbf{I}}}

\def\bY{{\mathbf{Y}}} \def\bZ{{\mathbf{Z}}}

\DeclarePairedDelimiter\norm{\lVert}{\rVert}

\newtheorem*{lemma*}{Lemma}

\begin{document}
	\title{Massive MIMO Channel Prediction Using Machine Learning: Power of Domain Transformation}
	
	
	\author{Beomsoo Ko,~\IEEEmembership{Student Member,~IEEE}, Hwanjin Kim,~\IEEEmembership{Graduate Student Member,~IEEE}, \\and~Junil Choi,~\IEEEmembership{Senior Member,~IEEE}
		\thanks{The authors are with the School of Electrical Engineering, Korea
			Advanced Institute of Science and Technology, Daejeon 34141, South Korea
			(e-mail: \{kobs0318, jin0903, junil\}@kaist.ac.kr).
			
		}
	}

	\maketitle
	

	\begin{abstract} \label{sec: absract} 
		To compensate the loss from outdated channel state information in wideband massive multiple-input multiple-output (MIMO) systems, channel prediction can be performed by leveraging the temporal correlation of wireless channels.
		Machine learning (ML)-based channel predictors for massive MIMO systems were designed recently; however, the time overhead to collect a large amount of training data directly affects the latency of the system.   
		In this paper, we propose a novel ML-based channel prediction technique, which can reduce the time overhead to collect the training data by transforming the domain of channels from subcarrier to antenna in wideband massive MIMO systems.  
		Numerical results show that the proposed technique can not only reduce the time overhead but also give additional performance gain compared to the ML-based channel prediction techniques without the domain transformation.
	\end{abstract}
	
	
	


	
	\section{Introduction}\label{sec1}
	Accurate channel state information is required in massive multiple-input multiple-output (MIMO) systems to fully exploit the advantage of a large number of antenna arrays \cite{Marzetta2010}.
	However, the estimated channel at the current coherence time block can be outdated due to the feedback delay of a system \cite{Ramya2009}, or the mobility of a user equipment (UE) \cite{Papa2017}.
	Channel prediction can address the outdated channel problem by using the estimated channels of the previous coherence time blocks without consuming additional pilot resources \cite{Kong2015}.
	
	Recently, machine learning (ML)-based channel predictors were developed using different types of neural networks \cite{Yuan2019, Kim2021, Jiang2019, Ko2021}. 
	A convolutional neural network (CNN) was combined with an autoregressive (AR) channel predictor in \cite{Yuan2019}, and a multi-layer perceptron (MLP)-based channel predictor was compared with the Kalman filter-based channel predictor in \cite{Kim2021}.
	A recurrent neural network (RNN) was employed for the channel prediction in wideband MIMO systems in \cite{Jiang2019}, where the frequency selective channel was transformed into multiple flat fading channels by using the orthogonal frequency division multiplexing (OFDM) technique.
	An MLP-based channel predictor was developed for wideband massive MIMO systems using OFDM in \cite{Ko2021}, where a single MLP is trained using the training data collected from a single subcarrier, and then predicts the channels of all subcarriers by exploiting the correlation of subcarrier channels.
	These works, however, did not pay much attention to reduce the time overhead of collecting the data to train neural networks, which should be minimized to make practical ML-based channel predictors.

	In this paper, we first demonstrate two naive ML-based channel prediction approaches for wideband MIMO systems using OFDM.
	The first approach separately generates ML-based channel predictors for each subcarrier channel, where the time overhead of collecting the training data is not considered.
	The second approach reduces the time overhead by training one neural network with the training data collected from every subcarrier channel.
	Although the second approach can reduce the time overhead to collect the same amount of training data as in the first approach, it turns out that high cross-correlation of the subcarrier channels in wideband massive MIMO degrades the quality of the training data, resulting in performance degradation. To overcome this issue, we propose a novel ML-based channel prediction technique that employs the domain transformation, i.e., transforms the subcarrier channels into the antenna-domain channels, to the second approach. The proposed technique can significantly reduce the time overhead to collect the training data and even have performance gain over the channel predictors without the domain transformation.
	
	The rest of paper is organized as follows. Section \ref{sec2} describes a system model, and Section \ref{sec3} demonstrates two naive ML-based channel prediction approaches for the wideband massive MIMO system. In Section \ref{sec4}, we propose a novel ML-based channel prediction technique that employs domain transformation. In Section \ref{simul}, we show numerical results of the ML-based channel prediction techniques. Finally, in Section \ref{conclusion}, we conclude our paper.
	
	\textbf{Notation:} 
	Column vectors and matrices are denoted with the lower and upper case of the bold letters, respectively. The multivariate complex normal distribution with mean vector $\boldsymbol{\mu}$ and covariance matrix $\boldsymbol{\Sigma}$ is represented as $\mathcal{CN}(\boldsymbol{\mu}, \boldsymbol{\Sigma})$.
	$\mathbf{0}_{m}$ denotes the $m \times 1$ all zero vector, and $\mathbf{I}_m$ is the $m \times m$ identity matrix.
	The Kronecker product is represented as the operator $\otimes$, and the vectorization of a matrix into a column vector is denoted as $\text{vec}(\cdot)$.
	The expectation is denoted as $\mathbb{E} \left[ \cdot \right]$, and the norm of a vector is denoted as $\lVert \cdot \rVert$. $\mathbb{C}^{m \times n}$ denotes the set of all $m \times n$ complex matrices. $\mathrm{Re}\left(a \right)$ and $\mathrm{Im}\left(a \right)$ denote the real and imaginary parts of a complex number $a$, respectively.

	\section{System Model And Channel Estimation}\label{sec2}
	\subsection{System Model}
	Since the channel estimation and prediction can be performed separately for each UE, we consider a single user wideband massive MIMO system in this paper.
	Due to the wide bandwidth of the system, the channel experiences frequency selectivity, which causes the inter-symbol-interference (ISI) problem. 
	To address this issue, we adopt the OFDM technique and transform the wideband channel into $L$ narrowband channels with individual subcarriers.  
	Assuming the base station (BS) has $M$ antennas, the uplink received signal of the $\ell$-th subcarrier at the $n$-th coherence time block is given as 
	\begin{align}\label{eq: eq1}
		\by^{\ell}_n=\sqrt{\rho}  \bh^{\ell}_{n} x^{\ell}_{n} + \bz^{\ell}_n,
	\end{align}
	where $\rho$ is the signal-to-noise ratio (SNR), $\bh^{\ell}_{n} \in \mathbb{C}^{M \times 1}$ is the uplink channel, $x^{\ell}_{n} \in \mathbb{C}$ is the transmit signal, and $\bz^{\ell}_n \sim \mathcal{CN}\left(\mathbf{0}_M, \bI_M \right)$ is the complex additive white Gaussian noise (AWGN). 
	
	\subsection{Channel Estimation}
	For the channel estimation, the UE transmits length $\tau$ pilot sequence to the BS.
	The overall received pilots of the $\ell$-th subcarrier at the $n$-th time block, $\bY^{\ell}_n$, is given as
	\begin{align}\label{eq: eq2}
		\bY^{\ell}_n = \sqrt{\rho} \bh^{\ell}_n  {\boldsymbol{\phi}^{\ell}_n}^{\mathrm{T}} + \bZ^{\ell}_n, 
	\end{align}
	where $\boldsymbol{\phi}^{\ell}_n \in \mathbb{C}^{\tau \times 1}$ is the length $\tau$ pilot sequence, and $\bZ^{\ell}_{n} \in \mathbb{C}^{M \times \tau}$ is the complex AWGN. 
	To estimate the channel from the received pilots, we transform (\ref{eq: eq2}) into 
	\begin{align}\label{eq: eq3}
		{\underline{\by}}^\ell_n = \boldsymbol{\Phi}^\ell_n \bh^\ell_n + \underline{\bz}^\ell_n,
	\end{align}
	where ${\underline{\by}}^\ell_n = \mathrm{vec}\left( \bY^{\ell}_n \right) \in \mathbb{C}^{M\tau \times 1}$, $\boldsymbol{\Phi}^\ell_n=\sqrt{\rho} {\boldsymbol{\phi}^{\ell}_n} \otimes \bI_M \in \mathbb{C}^{M\tau \times M}$, and ${\underline{\bz}}^\ell_n = \mathrm{vec}\left( \bZ^{\ell}_n \right) \in \mathbb{C}^{M\tau \times 1}$.
	The channel of the $\ell$-th subcarrier at the $n$-th time block can be estimated using the least square (LS) method, which is given as
	\begin{align}\label{eq: eq4}
		\bg_n^{\ell}  = \left( {\boldsymbol{\Phi}^\ell_n}^\mathrm{H} \boldsymbol{\Phi}^\ell_n \right)^{-1} {\boldsymbol{\Phi}^\ell_n}^\mathrm{H} \underline{\mathbf{y}}^{\ell}_n.
	\end{align}
	In our scenario of interest, the estimated channel $\bg_n^\ell$ becomes outdated. Hence, we develop ML-based channel prediction techniques that predict the true channel $\bh_{n+1}^\ell$ using $\left\{\bg_i^\ell\right\}_{i \leq n}$. We denote the predicted channel of the $\ell$-th subcarrier at the $n$-th coherence time block as $\hat{\bh}_{n+1}^\ell$.

	\section{Separate And Joint Learning Approaches}\label{sec3}
	In this section, we demonstrate two naive ML-based wideband massive MIMO channel prediction approaches. The first approach is a separate learning (SL) approach, where the channel predictors are generated separately for each subcarrier channel. The second approach is a joint learning (JL) approach, where the channel predictor is generated by training a single neural network with the training data that are collected from every subcarrier channel.
	
	\subsection{SL Approach}
	The SL approach simply repeats the ML-based channel prediction technique in \cite{Kim2021} for each subcarrier channel. 
	The optimization problem for the $\ell$-th subcarrier is given as
	\begin{align}
		& \text{min} \  \mathbb{E} \left[\norm{ \bg^{\ell}_{n+1} - {\hat{\bh}}^{\ell}_{n+1} }^2 \right], \label{eq: eq5} \\
		& \text{s.t.} \  {\hat{\bh}}^{\ell}_{n+1} = f^\ell \left( {{\bg}}^\ell_{n-n_0+1}, \cdots , {{\bg}}^\ell_n \right), \label{eq: eq6}
	\end{align}
	where the objective is to find the channel predictor $f^\ell(\cdot)$ that minimizes the mean square error (MSE) between the estimated channel $\bg^{\ell}_{n+1}$ and the predicted channel ${\hat{\bh}}^{\ell}_{n+1}$. The input order $n_0$ is a control parameter, which is related to the channel variation in time \cite{Kim2021}. The estimated channel $\bg^{\ell}_{n+1}$ is used in the optimization process since the true channel $\bh_{n+1}^\ell$ cannot be measured in practice.
	\begin{figure}
		\centering
		\includegraphics[width=0.90\columnwidth]{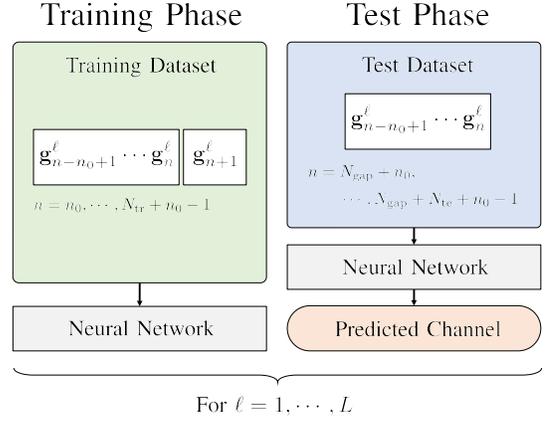}
		\caption{Two-phase block diagram of the SL approach.}\label{fig: SDSL}
	\end{figure}

	The optimization problem can be solved via training the neural network with the training data.
	The training and test datasets for the $\ell$-th subcarrier are denoted as
	\begin{align}
		\mathcal{D}^\ell_\mathrm{tr}(N_\mathrm{tr}) & =\left( \left\{{{\bg}}^\ell_{n-n_0+1}, \cdots , {{\bg}}^\ell_n \right\}, {{\bg}}^\ell_{n+1} \right)_{n=n_0}^{N_{\mathrm{tr}}+n_0-1}, \nonumber  \\
		\mathcal{D}^\ell_\mathrm{te}(N_\mathrm{te}) & =\left( \left\{{{\bg}}^\ell_{n-n_0+1}, \cdots , {{\bg}}^\ell_{n} \right\} \right)_{n=N_\mathrm{gap}+n_0}^{N_\mathrm{gap}+N_{\mathrm{te}}+n_0-1}, \label{eq: eq7}
	\end{align}
	where $N_\mathrm{tr}$ and $N_\mathrm{te}$ are the total numbers of training and test data collected from each subcarrier, respectively. To separate the training data and test data, $N_{\mathrm{gap}}$ satisfies $N_\mathrm{tr}<N_{\mathrm{gap}}$. The features of the training data are denoted as $\left\{{{\bg}}^\ell_{n-n_0+1}, \cdots , {{\bg}}^\ell_n \right\}$, and the label is denoted as ${{\bg}}^\ell_{n+1}$. 
	The detail of training and test phases of the SL approach is described in Fig. \ref{fig: SDSL}. In the training phase, the training dataset $\mathcal{D}^\ell_\mathrm{tr}(N_\mathrm{tr})$ is fed into the neural network. After training the neural network, the channel predictor is generated.
	Then, the test dataset $\mathcal{D}^\ell_\mathrm{te}(N_\mathrm{te})$ is fed into the channel predictor to predict the channel of the $\ell$-th subcarrier. 
	
	\subsection{JL Approach}
	Although the SL approach can be easily applied to the wideband massive MIMO system, the time overhead to collect a sufficient amount of training data should be also considered.
	In the JL approach, a single channel predictor is generated by training one neural network with the training data collected from every subcarrier to reduce the time overhead. The JL approach is motivated from the ML-based channel prediction technique in \cite{Ko2021}, which exploits the correlation of subcarrier channels. 
	
	The optimization problem for the JL approach is given as
	\begin{align}
		& \text{min} \  \sum_{\ell=1}^L \mathbb{E} \left[\norm{ \bg^{\ell}_{n+1} - {\hat{\bh}}^{\ell}_{n+1} }^2 \right], \label{eq: eq8}\\
		& \text{s.t.} \  {\hat{\bh}}^{\ell}_{n+1} = f \left( {{\bg}}^\ell_{n-n_0+1}, \cdots , {{\bg}}^\ell_n \right),\forall \ell=1, \cdots, L,\label{eq: eq9}
	\end{align}
	where the objective is to generate a single channel predictor $f(\cdot)$ for every subcarrier.
	The main difference from the SL approach is that the MSE is computed over all subcarriers.
	\begin{figure}
		\centering
		\includegraphics[width=0.90\columnwidth]{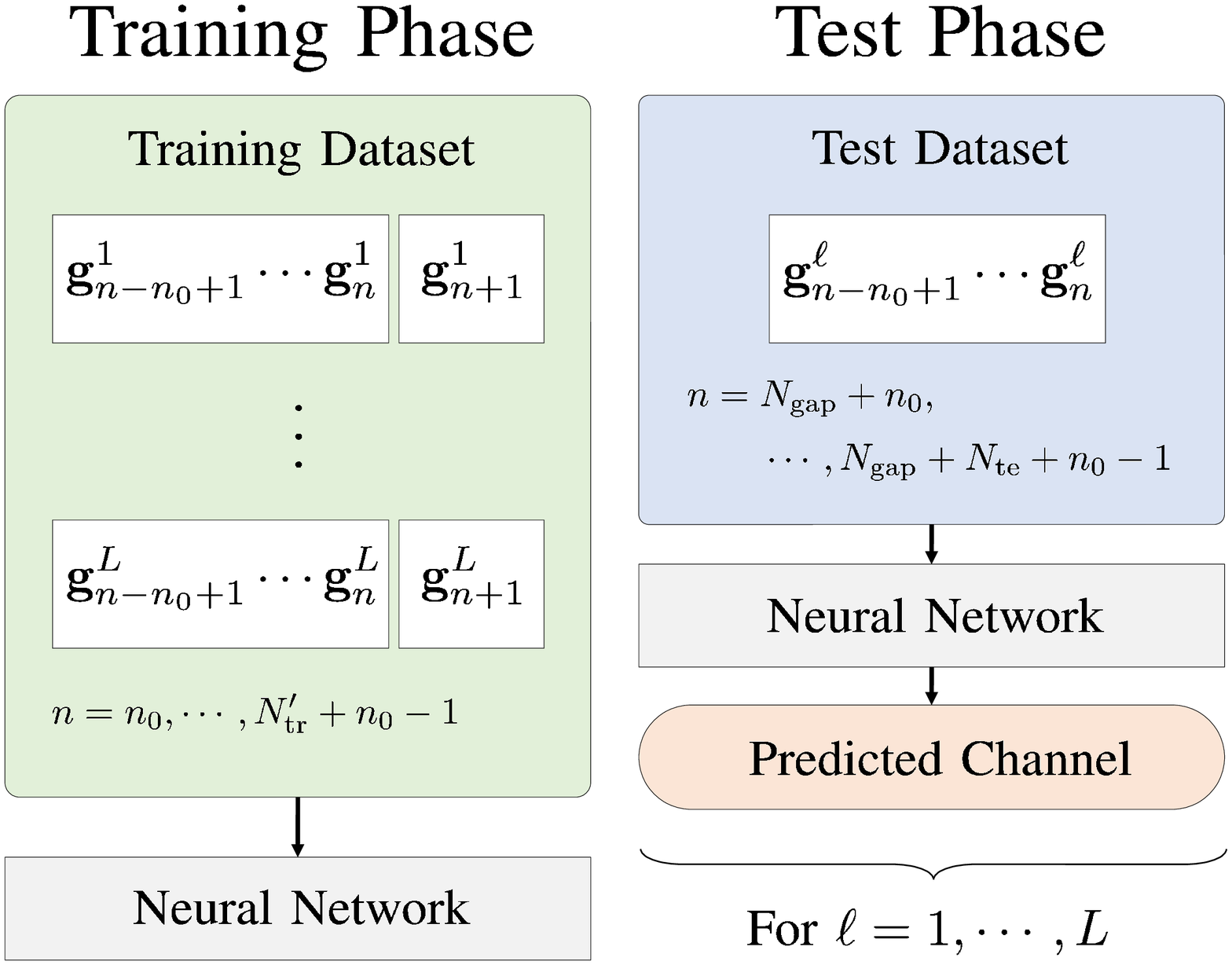}
		\caption{Two-phase block diagram of the JL approach.}\label{fig: SDJL}
	\end{figure}
	The training and test datasets for the JL approach are given as
	\begin{align}
		\mathcal{D}_\mathrm{tr}^\mathrm{JL} &=\bigcup_{\ell=1}^L \mathcal{D}_{\mathrm{tr}}^{\ell}(N_\mathrm{tr}^\prime), \nonumber\\
		\mathcal{D}_\mathrm{te}^\mathrm{JL} &=\bigcup_{\ell=1}^L \mathcal{D}_{\mathrm{te}}^{\ell}(N_\mathrm{te}),\label{eq: eq10}
	\end{align}
	where $N_\mathrm{tr}^\prime$ is the number of the training data collected from each subcarrier. To make fair comparison with the SL approach, we set $N_\mathrm{tr}^\prime=N_\mathrm{tr}/L$; therefore, the time overhead for collecting the training data from each subcarrier is reduced with the number of subcarriers $L$ compared to the SL approach.
	The detail of the JL approach is described in Fig. \ref{fig: SDJL}.
	In the training phase, the training dataset $\mathcal{D}_\mathrm{tr}^\mathrm{JL}$ is fed into the neural network. After the training, the generated channel predictor is used to predict the channel for each subcarrier.
	
		\textit{Remark:}
		While the time overhead to collect the training data can be reduced linearly with $L$ for each subcarrier in the JL approach, the channel prediction performance is not guaranteed since the correlation properties of the training data can affect the performance of the ML-based channel predictors.
		Although the JL approach is to exploit the correlation of subcarrier channels, it is shown in Section \ref{sec4}-B that the cross-correlation of subcarrier channels is too high, which makes the training dataset highly redundant, resulting in performance degradation \cite{Biro2019}. Therefore, we investigate the way to reduce the correlation of the training data in next the section.

	\section{Domain Transformation}\label{sec4}
	In the wideband massive MIMO system, it is natural to express the channel in the subcarrier-domain.
	However, we are motivated to express the channel in the antenna-domain using domain transformation since the collected training data from all subcarrier channels are highly redundant due to the large cross-correlation of the subcarrier channels. 
	In this section, we first introduce the domain transformation of the wideband massive MIMO system, which transforms the subcarrier channels into the antenna-domain channels. 
	Next, we investigate the auto- and cross-correlations of both the subcarrier channels and antenna-domain channels, where the auto-correlations of the channels reflect the correlation between the features of the training data, and the cross-correlations of the channels reflect the correlation between the training datasets, respectively.
	Finally, we propose an ML-based channel predictor that employs the domain transformation to the JL approach, which reduces the time overhead of collecting the training data and gives additional prediction performance gain compared to the SL and JL approaches.
	
	\subsection{Domain Transformation}
	To explain the domain transformation, we first denote the $m$-th element of $\bh_n^\ell$, $\bg_n^\ell$, and $\hat{\bh}_{n+1}^\ell$ as $h^{\ell, m}_n$, $g^{\ell, m}_n$, and $\hat{h}^{\ell ,m}_{n+1}$, respectively.
	We then define the true, estimated, and predicted channels of the $m$-th antenna at the $n$-th coherence time block as
	\begin{align}
		\mathbf{\mathfrak{h}}_n^m &= \left[h^{1 ,m}_n, \cdots, h^{L, m}_n \right]^\mathrm{T}, \label{eq: eq11} \\
		\mathbf{\mathfrak{g}}_n^m &= \left[g^{1 ,m}_n, \cdots, g^{L, m}_n \right]^\mathrm{T}, \label{eq: eq12} \\
		\hat{\mathbf{\mathfrak{h}}}_{n+1}^m &= \left[\hat{h}^{1 ,m}_{n+1}, \cdots, \hat{h}^{L, m}_{n+1} \right]^\mathrm{T}, \label{eq: eq13}
	\end{align}
	where $\left(\mathbf{\mathfrak{h}}^m_n\right)^\mathrm{T} \in \mathbb{C}^{1 \times L}$ is the row vector of the matrix $\left[\bh_n^{1}, \cdots,\bh_n^{L} \right] = \left[\mathbf{\mathfrak{h}}^1_n, \cdots, \mathbf{\mathfrak{h}}^M_n \right]^\mathrm{T}$. Note that the elements of  $\mathbf{\mathfrak{h}}^m_n$ are the subcarrier channels that the $m$-th antenna sees. Since the expressions in (\ref{eq: eq11})-(\ref{eq: eq13}) are for the $m$-th antenna, we call these channels as the antenna-domain channels. 
	
	\subsection{Correlation Properties}
	Before applying the domain transformation to the ML-based channel prediction, we investigate the auto- and cross-correlations of both the subcarrier channels and antenna-domain channels. 
	The auto-correlation of the $\ell$-th subcarrier channel can be computed as
	\begin{align}\label{eq: eq14}
		R_{\ell}(\tau)= \frac{1}{N_\mathrm{avg}} \sum_{n=1}^{N_\mathrm{avg}} {\bh_n^\ell}^\mathrm{H} \bh_{n+\tau}^{\ell},
	\end{align}
	where $\tau$ is the time shift, and $N_\mathrm{avg}$ is the total number of samples for the sample average.
	\begin{figure}
		\centering
		\includegraphics[width=0.90\columnwidth]{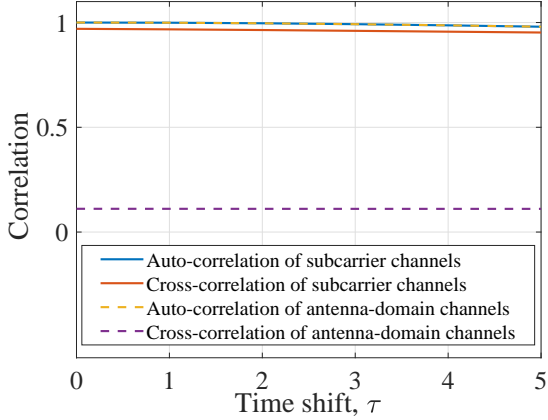}
		\caption{Auto- and cross-correlation of subcarrier channels and antenna-domain channels.}\label{fig: Corr}
	\end{figure}
 	The cross-correlation between the $\ell$-th and $\ell^\prime$-th subcarrier channels can be computed as
	\begin{align}\label{eq: eq15}
		R_{\ell, \ell^\prime}(\tau)=\frac{1}{N_\mathrm{avg}} \sum_{n=1}^{N_\mathrm{avg}} {\mathbf{h}_n^\ell }^\mathrm{H} \mathbf{{h}}_{n+\tau}^{\ell^\prime}.
	\end{align}
	The auto- and cross-correlations of the antenna-domain channels can be also computed similarly with the channels defined in (\ref{eq: eq11}). 

	In Fig. \ref{fig: Corr}, we investigate the correlation properties of both the subcarrier channels and antenna-domain channels with respect to the time shift $\tau$ considering the channels used in Section \ref{simul}.
	The total number of samples $N_\mathrm{avg}$ is set to 2000.
	The correlation values for the subcarrier channels and antenna-domain channels are averaged out for every subcarrier and antenna, respectively.
	We first observe that the auto-correlation values of the subcarrier channels and antenna-domain channels are both high regardless of $\tau$. These conditions are beneficial when predicting the channel based on the previous channel estimates because the channels are highly correlated in time.
	We also observe that the cross-correlation values of the subcarrier channels are close to 1 for all $\tau$, which reflects the high redundancy of the training dataset in the JL approach.
	Since the cross-correlation values of the antenna-domain channels are close to 0, we are motivated to use the domain transformation from the subcarrier-domain to the antenna-domain to reduce the redundancy of the training data in the JL approach.	

	\subsection{JL With The Domain Transformation}
	Based on the discussion of correlation properties in the previous subsection, we employ the domain transformation to the JL approach. We denote the proposed technique as a joint learning with domain transformation (JLDT) approach for the rest of the paper.
	The subcarrier channels are transformed into the antenna-domain channels, and then the neural network is trained with every antenna-domain channel. 
	The JLDT approach shares the same motivation with the JL approach, which is to reduce the time overhead from collecting the training data. Hence, we use the exact same amount of time overhead to collect the training data from each subcarrier channel as in the JL approach.
	The optimization problem for the JLDT approach is given as
	\begin{align}
		&\text{min} \  \sum_{m=1}^M \mathbb{E} \left[\norm{ \mathbf{\mathfrak{g}}^{m}_{n+1} - {\hat{\mathbf{\mathfrak{h}}}}^{m}_{n+1} }^2 \right],\label{eq: eq17} \\
		&\text{s.t.} \  {\hat{\mathbf{\mathfrak{h}}}}^{m}_{n+1} = f \left( {{\mathbf{\mathfrak{g}}}}^m_{n-n_0+1}, \cdots , {{\mathbf{\mathfrak{g}}}}^m_n \right),\forall m=1, \cdots, M, \label{eq: eq18}
	\end{align}
	where the channel vectors are in the form of the antenna-domain described in (\ref{eq: eq12}) and (\ref{eq: eq13}), and the objective is to generate a single channel predictor $f(\cdot)$ for every antenna-domain channel.
	The training and test datasets for the JLDT approach are given as
	\begin{align}
		\mathcal{D}_\mathrm{tr}^\mathrm{JLDT} &=\bigcup_{m=1}^M \left( \left\{{{{\mathbf{\mathfrak{g}}}}}^m_{n-n_0+1}, \cdots , {{{\mathbf{\mathfrak{g}}}}}^m_n \right\}, {{{\mathbf{\mathfrak{g}}}}}^m_{n+1} \right)_{n=n_0}^{N_{\mathrm{tr}}^\prime+n_0-1}, \nonumber \\
		\mathcal{D}_\mathrm{te}^\mathrm{JLDT} &=\bigcup_{m=1}^M \left( \left\{{{{\mathbf{\mathfrak{g}}}}}^m_{n-n_0+1}, \cdots , {{{\mathbf{\mathfrak{g}}}}}^m_{n} \right\} \right)_{n=N_\mathrm{gap}+n_0}^{N_\mathrm{gap}+N_{\mathrm{te}}+n_0-1}.
	\end{align}
	
	Although the training data in the JLDT approach have different form compared to the JL approach, both approaches collect $N_\mathrm{tr}^\prime$ amount of training data from each subcarrier channel and antenna-domain channel, respectively.  
	Hence, the time overhead for collecting the training data in the JLDT approach is reduced linearly by $L$ as in the JL approach.
	In the training phase, the training dataset $\mathcal{D}_\mathrm{tr}^\mathrm{JLDT}$ is fed into the neural network. After the training, the generated channel predictor is used to predict the channel of every antenna. 
	From the predicted antenna-domain channels, we reconstruct the subcarrier channels.

	\section{Numerical Results}\label{simul}
	\begin{figure}
		\centering
		\includegraphics[width=0.90\columnwidth]{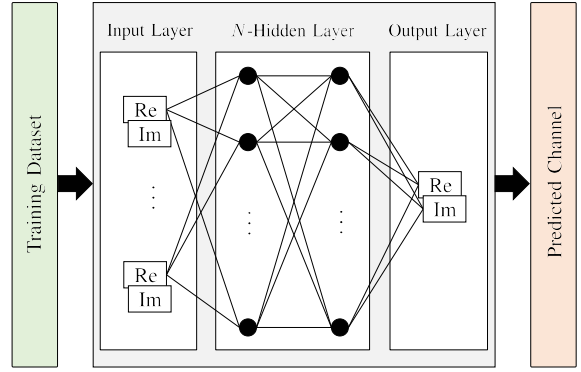}
		\caption{MLP structure with input layer, $N$ hidden layers, and output layer.}\label{fig: mlp}
	\end{figure}
	
	In this paper, an MLP is implemented for the neural network. 
	As described in Fig. \ref{fig: mlp}, the MLP includes the input layer, $N$ hidden layers, and output layer.
	For the SL approach, the input data has the form of $\left\{{{\bg}}^\ell_{n-n_0+1}, \cdots , {{\bg}}^\ell_n \right\}$, which are the estimated subcarrier channels of the previous time blocks. In the MLP, the input data is split into the real and imaginary parts, i.e., $\left\{ \mathrm{Re} \left(  {{\bg}}^\ell_{n-n_0+1}\right), \mathrm{Im} \left( {{\bg}}^\ell_{n-n_0+1} \right),  \cdots , \mathrm{Re} \left( {{\bg}}^\ell_n \right), \mathrm{Im} \left( {{\bg}}^\ell_n \right)  \right\}$, since the input should be real valued. The input data is then vectorized so that the dimension of the input layer and the output layer are $2n_0M$ and $2M$, respectively. Similarly, since the output of the MLP are real valued, we recombine the output values into the complex channel, which corresponds to the predicted channel ${\hat{\bh}}^{\ell}_{n+1}$. The input and output data for the JL and JLDT approaches are split and recombined in a similar manner with the SL approach.
	For the optimizer, an adaptive moment estimation (ADAM) in \cite{Kingma2015} is employed, and the cost function is the MSE between the estimated channel $\bg^{\ell}_{n+1}$ and the predicted channel ${\hat{\bh}}^{\ell}_{n+1}$. 
	
	In this paper, the channels are generated using the quasi deterministic radio channel generator (QuaDRiGa) in \cite{QUA}. The urban micro (UMi) scenario is considered, where the mobility of UE is set to 1 $\mathrm{km/h}$.
	The coherence time block duration is 20 $\mathrm{ms}$, the carrier frequency is 2.53 $\mathrm{GHz}$, and the number of subcarriers is 50. We implement a uniform planar array (UPA) of 8 by 8 antennas at the BS.   
	For the MLP, two hidden layers with 512 nodes are utilized. The input order $n_0$ is set to 3.
	We set the size of the batch to 128, learning rate to 0.001, and epoch to 1000. 
	The discrete Fourier transform (DFT) matrix is used for the pilot $\boldsymbol{\Phi}^\ell_n$ in the channel estimation.
	The performance of the channel prediction is computed with the normalized MSE (NMSE), which is defined as
	\begin{align}
		\mathrm{NMSE}=\mathbb{E} \left[ \norm{ \bh^{\ell}_{n+1} - {\hat{\bh}}^{\ell}_{n+1} }^2 /  \norm{{{\bh}}^{\ell}_{n+1} }^2  \right].
	\end{align}
	
	\begin{figure}
		\centering
		\includegraphics[width=0.90\columnwidth]{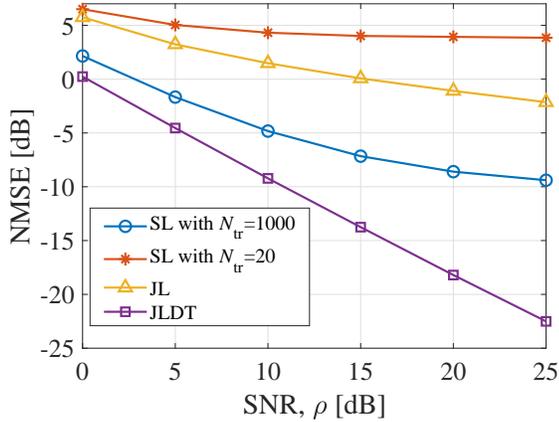}
		\caption{NMSE performance of the SL, JL and JLDT approaches with respect to the SNR.}\label{fig: nmse}
	\end{figure}

	In Fig. \ref{fig: nmse}, we investigate the performance of the channel prediction of the SL approach, JL approach, and JLDT approach with respect to the SNR. The number of training data collected from each subcarrier for the JL and JLDT approaches is $N_\mathrm{tr}^\prime=20$. We also investigate the performance of the SL approach with $N_\mathrm{tr}=1000$ and $N_\mathrm{tr}=20$. The case of $N_\mathrm{tr}=1000$ is to match the total number of training data with the JL approach, and the case of $N_\mathrm{tr}=20$ is to have the same time overhead with the JL and JLDT approaches. 
	
	The figure shows that the SL approach with $N_\mathrm{tr}=20$ has the worst prediction performance since the number of training data is not sufficient to train the neural network.
	The JL approach experiences significant performance loss compared to the SL approach with  $N_\mathrm{tr}=1000$. The high cross-correlation of the subcarrier channels led to the over-fitting in the JL approach since the training data that are collected from different subcarrier channels are highly correlated.
	On the contrary, the JLDT approach has a performance gain compared to the SL approach with $N_\mathrm{tr}=1000$ while consuming the same amount of the time overhead with the JL approach. This gain comes from the low cross-correlation of the antenna-domain channels as shown in Fig. \ref{fig: Corr}, making all the training data for the JLDT approach highly effective.

	\section{Conclusion}\label{conclusion}
	This paper proposed a novel ML-based channel predictor for the massive MIMO wideband system using OFDM.
	We first introduced two naive approaches to generate ML-based channel predictors, which are the SL and JL approaches. 
	To reduce the large time overhead of the SL approach, the JL approach and JLDT approach train a single neural network with every subcarrier and antenna-domain channel, respectively. Different from the JL approach, the JLDT approach adopts the domain transformation, which transforms the subcarrier channels into the antenna-domian channels, on the training data.
	The numerical results showed the JLDT approach had superior channel prediction performance over both the SL approach and JL approach despite of its very small time overhead to collect the training data.
	The low cross-correlation of antenna-domain channels in the JLDT approach is beneficial to enhance the quality of the training data. 
	We believe the concept of domain transformation, i.e., transform the domain of training data to have low cross-correlation, can reduce the time overhead to collect the training data and enhance the performance of many ML-based wireless communication systems.

	\bibliographystyle{IEEEtran}
	\bibliography{Reference}
	
\end{document}